\documentclass[twocolumn,superscriptaddress,aps,prl]{revtex4-1}
\usepackage{amsfonts}
\usepackage{amsmath}
\usepackage{amssymb}
\usepackage{graphicx}
\usepackage{color}
\usepackage{ulem}
\usepackage{dcolumn}
\usepackage{bm}
\usepackage{url}
\usepackage[colorlinks, urlcolor = cyan, citecolor = blue, linkcolor = magenta]{hyperref}
\usepackage{ulem}

\begin{document}

\title{Floquet Dynamical Decoupling at Zero Bias}
\author{Peng Xu}
\email{2014202020003@whu.edu.cn}
\affiliation{School of Physics, Zhengzhou University, Zhengzhou 450001, China}
\author{Jun Zhang}
\affiliation{School of Physics and Technology, Wuhan University, Wuhan, Hubei 430072, China}
\date{\today}

\begin{abstract}
Dynamical decoupling (DD) is an efficient method to decouple systems from environmental noises and to prolong the coherence time of systems. In contrast to discrete and continuous DD protocols in the presence of bias field, we propose a Floquet DD at zero bias to perfectly suppress both the zeroth and first orders of noises according to the Floquet theory. Specifically, we demonstrate the effectiveness of this Floquet DD protocol in two typical systems including a spinor atomic Bose-Einstein condensate decohered by classical stray magnetic fields and a semiconductor quantum dot electron spin coupled to nuclear spins. Furthermore, our protocol can be used to sense high-frequency noises. The Floquet DD protocol we propose shines new light on low-cost and high-portable DD technics without bias field and with low controlling power, which may have wide applications in quantum computing, quantum sensing, nuclear magnetic resonance and magnetic resonance imaging.
\end{abstract}

\maketitle

{\it{Introduction.}} Systems that are not sufficiently isolated inevitably couple to environments, resulting in finite coherence time, finite lifetime and particle loss~\cite{Nielsen2002Quantum, Breuer2002Theory, Gardiner2004Quantum, Suter2016Protecting}. The fidelity of entangled state preparations and the reliability of quantum gate operations will be decreased~\cite{Ma2011Quantum, Leandro2015Open, Xu2019Efficient, Knill2005Quantum, Abdelhafez2020Universal}, leading to destructive effects on many quantum applications, such as quantum communication~\cite{Bouchard2021Achieving}, quantum teleportation~\cite{Oh2002Fidelity, Jung2008Greenberger}, and quantum computing~\cite{Aliferis2008Fault, Urbanek2021Mitigating}. Therefore, it is crucial to suppress these destructive channels in experiments. However, the timescale of decoherence induced by noises or interactions is much shorter than other destructive channels. Thus suppressing the decoherence channel and prolonging the coherence time is the first challenge in experiments~\cite{Paik2008Decoherence, Pokharel2018Demonstration, Abobeih2018One, Bauch2020Decoherence, Wang2017Single, Wang2021Single}. 

Dynamical decoupling (DD) is a useful mechanism to prolong the coherence time of systems and to decouple systems from both classical and quantum noises~\cite{Haeberlen2012High, Mehring2012Principles, Slichter2013Principles}. It was originally proposed by Hahn as spin echo~\cite{Hahn1950Spin}, then developed into various forms, such as Carr-Purcell-Meiboom-Gill (CPMG)~\cite{Carr1954Effects, Meiboom1958Modified}, periodic DD (PDD)~\cite{Khodjasteh2005Fault}, concatenated DD (CDD)~\cite{Khodjasteh2005Fault, Zhang2008Long}, Uhrig DD (UDD)~\cite{Uhrig2007Keeping} and uniaxial DD (Uni-DD)~\cite{Yao2019Uniaxial}. The DD protocols mentioned above are the discrete type with strong strength of control pulses. In order to decrease the controlling power, continuous DD protocols have been proposed~\cite{Fanchini2007Continuously1, Fanchini2007Continuously2, Bermudez2012Robust, Chaudhry2012Decoherence, Cai2012Robust, Timoney2011Quantum, Xu2012Coherence, Zhang2016Preserving, Stark2017Narrow, Anderson2018Continuously, Cai2022Optimizing}. Recently, these DD protocols are widely applied to various systems, such as quantum dots~\cite{Medford2012Scaling, Sun2022Full}, nitrogen-vacancy centers~\cite{Xu2012Coherence, Anderson2018Continuously, Abobeih2018One}, trapped ions~\cite{Timoney2011Quantum, Wang2017Single, Wang2021Single} and superconducting quantum interference devices~\cite{Guo2018Dephasing, Pokharel2018Demonstration, Qiu2021Suppressing}. The coherence time of systems is significantly improved in experiments. However, both discrete and continuous DD protocols require a large bias field, which makes the experimental setups high-cost and low-portable. Besides, there are potential tasks to decouple and to sense noises without bias~\cite{Dmitriev2018Multi, Dmitriev2019Dual, Dmitriev2022Radio, Mikawa2023Electron}. So, it is necessary and significant to propose a novel protocol aiming at suppressing the decoherence channel and sensing high-frequency noises at zero bias.

In this letter, we propose a new method named Floquet DD at zero bias to decouple systems from environments. We first analytically obtain the effective Hamiltonian under Floquet DD based on the Floquet theory, which demonstrates our protocol can not only suppress the zeroth order of noises but prevent systems from being decohered by the first order of noises. Second, numerical simulations confirm the commendable performance of our protocol in a spinor Bose-Einstein condensate (BEC) decohered by stray magnetic fields (classical noises) and in a semiconductor quantum dot (QD) electron spin qubit decohered by nuclear spins (quantum noises). Our results can be applied to research in low control field and zero bias in nuclear magnetic resonance and quantum sensing beyond standard quantum limit.

\begin{figure}[t]
    \centering
    \includegraphics[width=7.8cm]{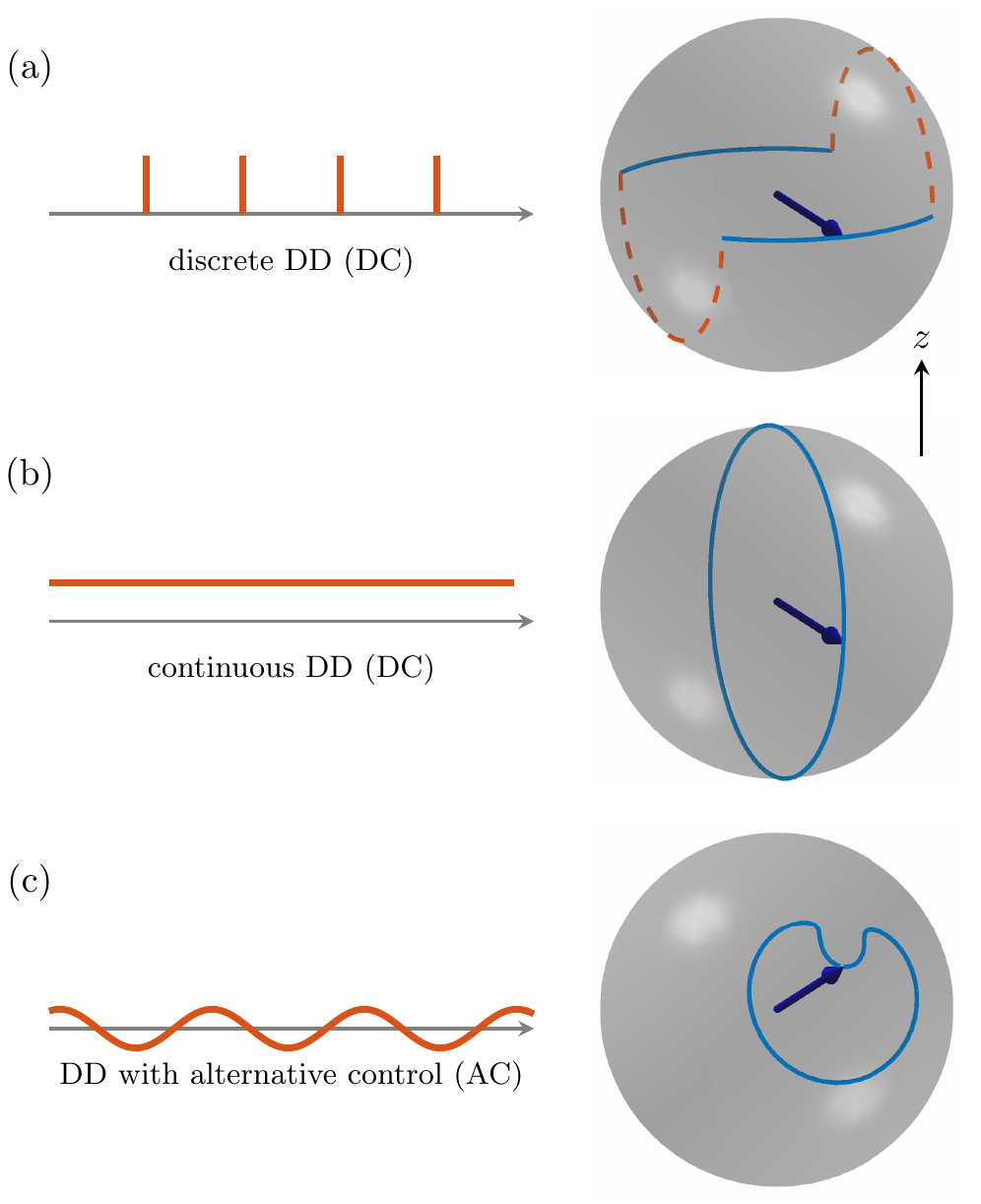}
    \caption{Schematics for different DD protocols. The control is along $x$-axis, and noises are along $z$-axis, $b_z \in [- b_m, b_m]$. The typical DD protocols are shown in the left panels, while the corresponding evolutions are shown in the right panels. (a) Discrete DD. (b) Continuous DD. (c) Floquet DD.}
    \label{fig:setup}
\end{figure}

{\it{Floquet DD.}} Before we discuss the Floquet DD, we briefly review existing DD protocols. To compare our method with these protocols, we depict them in the toggling frame of bias since there is no bias in our method intrinsically. The simplest discrete DD, i.e., spin echo, is shown in Fig.~\ref{fig:setup} (a). The control Hamiltonian is $\hat{H}_c = \Omega \delta(t - nT) \hat{J}_x$ with controlling strength satisfying $\int_{nT - \epsilon / 2}^{nT + \epsilon / 2} \Omega \delta(t - nT) dt = \pi$, then noises along $z$-axis $\hat{H}_z = b_z \hat{J}_z$ with strength $b_z \in [- b_m, b_m]$ are suppressed. As a result, the spin returns to its initial state after one period as shown in the right panel of Fig.~\ref{fig:setup} (a). A typical continuous DD is shown in Fig.~\ref{fig:setup} (b). The control Hamiltonian is $\hat{H}_c = \Omega \hat{J}_x$, then the system is decoupled from noises under the condition $\Omega \gg b_m$. The evolution trajectory is shown in the right panel of Fig.~\ref{fig:setup} (b), which forms a closed circle. 

The Floquet DD protocol we propose is shown in Fig.~\ref{fig:setup} (c). The control Hamiltonian is $\hat{H} = \Omega \cos(\omega t + \varphi) \hat{J}_x$ with controlling strength $\Omega$ and frequency $\omega$, then the evolution of systems coupled to longitudinal magnetic noises is governed by 
\begin{equation}
    \hat{H} = b_z \hat{J}_z + \Omega \cos(\omega t + \varphi) \hat{J}_x,
\end{equation}
where we assume $|b_z| \leq b_m$. After applying a rotating wave transformation $\hat{U} = e^{i \int_0^t \Omega \cos(\omega \tau + \varphi) \hat{J}_x d\tau}$, the Hamiltonian becomes
\begin{equation}
    \begin{aligned}
    \hat{H}_r = b_z &\sum_{n = - \infty}^{\infty} \left[ \mathcal{J}_{2 n} e^{i 2 n (\omega t + \varphi)} \hat{J}_z - i \mathcal{J}_{2 n + 1} e^{i (2 n + 1) (\omega t + \varphi)} \hat{J}_y \right],
    \end{aligned}
\end{equation}
where $\mathcal{J}_n$ is the $n$-th order Bessel function of $\Omega / \omega$. When $\omega$ is larger than $b_m$, we can use the high-frequency expansion to obtain the Floquet effective Hamiltonian, $\hat{H}_{\text{eff}} = \hat{H}^{(0)} + \sum_{n \neq 0} \left( \frac{ \left[ \hat{H}^{(n)}, \hat{H}^{(- n)} \right]}{2 n \omega} - \frac{\left[ \hat{H}^{(n)}, \hat{H}^{(0)} \right]}{n \omega} \right) + \cdots$, where $\hat{H}^{(n)} = \frac{1}{T} \int_{0}^{T} \hat{H}(t) e^{- i n \omega t} dt$. In the high-frequency limit, the rotating wave approximation, we can keep the expansion to the leading order
\begin{equation}
    \hat{H}_{\text{eff}} = b_z \mathcal{J}_{0} \hat{J}_z.
\end{equation}
When the ratio $\Omega / \omega$ satisfies $\mathcal{J}_0 \left( \Omega / \omega \right) = 0$, the zeroth order of noises about $b_z$ is completely suppressed. Furthermore, because of $\left[\hat{H}^{(n)}, \hat{H}^{(- n)}\right] = 0$, the first order of longitudinal noises is also inhibited perfectly.

According to the above analysis, we conclude that both the zeroth and first orders of longitudinal noises are suppressed completely. However, in experiments, there may exist noises with random directions, i.e., stray magnetic fields. So, it is vital for Floquet DD to prevent systems from being decohered by stray magnetic fields. Fortunately, our protocol can be directly extended to suppress these noises. The Hamiltonian with controls in this situation is
\begin{equation}
    \begin{aligned}
    \hat{H} = \bm{b \cdot \hat{J}} + \sum_{\alpha = x, y} \Omega_\alpha \cos(\omega_\alpha t + \varphi_\alpha) \hat{J}_\alpha, 
    \end{aligned}
    \label{eq:DD_Cl_H}
\end{equation}
where we assume $|\bm{b}| \leq b_m$ and $\omega_x = \omega_y = \omega$~\cite{note2}. In the following, we apply a rotating wave transformation $\hat{U}^{\text{I}} = e^{i \int_0^t \Omega_x \cos(\omega \tau + \varphi_x) \hat{J}_x d\tau}$, resulting in the Hamiltonian
\begin{equation}
    \hat{H}_r^{\text{I}} = \sum_{n = -\infty}^{\infty} \hat{H}^{(n)} e^{i n \omega t},
    \label{eq:DD_Cl_H_r1}
\end{equation}
where
\begin{equation}
    \begin{aligned}
    &\hat{H}^{(0)} = b_x \hat{J}_x + b_y \mathcal{J}_{0} \hat{J}_y + b_z \mathcal{J}_{0} \hat{J}_z - \Omega_y \mathcal{J}_{1} \sin(\varphi_x - \varphi_y) \hat{J}_z, \\
    &\hat{H}^{(2 n \neq 0)} =  b_y \mathcal{J}_{2 n} e^{i 2 n \varphi_x} \hat{J}_y + b_z \mathcal{J}_{2 n} e^{i 2 n \varphi_x} \hat{J}_z \\ 
    &+ i \frac{\Omega_y}{2} \left[ \mathcal{J}_{2 n - 1} e^{i [(2 n - 1) \varphi_x + \varphi_y]} + \mathcal{J}_{2 n + 1} e^{i [(2 n + 1) \varphi_x - \varphi_y]} \right] \hat{J}_z \\
    &\hat{H}^{(2 n + 1)} = i b_y \mathcal{J}_{2 n + 1} e^{i (2 n + 1) \varphi_x} \hat{J}_z - i b_z \mathcal{J}_{2 n + 1} e^{i (2 n + 1) \varphi_x} \hat{J}_y \\
    &+ \frac{\Omega_y}{2} \left[ \mathcal{J}_{2 n} e^{i [2 n \varphi_x + \varphi_y]} + \mathcal{J}_{2 n + 2} e^{i [(2 n + 2) \varphi_x - \varphi_y]} \right] \hat{J}_y, \nonumber
    \end{aligned}
\end{equation}
with $\mathcal{J}_n \equiv \mathcal{J}_n\left( \Omega_x / \omega \right)$. In the high-frequency limit $\omega \gg b_m, \Omega_y$ and $\mathcal{J}_0 \left(\Omega_x / \omega\right) = 0$, the effective Hamiltonian becomes
\begin{equation}
    \hat{H}_{\text{eff}}^{\text{I}} = b_x \hat{J}_x + \gamma \hat{J}_z,
    \label{eq:DD_Cl_H_eff1}
\end{equation}
with $\gamma = - \Omega_y \mathcal{J}_{1} \sin(\varphi_x - \varphi_y)$. Then we apply another rotating wave transformation $\hat{U}^{\text{II}} = e^{i \gamma \hat{J}_z t}$, which transforms the above Hamiltonian to
\begin{equation}
    \hat{H}_r^{\text{II}} = b_x \left[ \cos(\gamma t) \hat{J}_x - \sin(\gamma t) \hat{J}_y \right].
\end{equation}
In the rotating wave approximation $\gamma \gg b_m$, the effective Hamiltonian is finally obtained 
\begin{equation}
    \hat{H}_{\text{eff}}^{\text{II}} = 0.
\end{equation}
Thus, the zeroth order of stray magnetic fields is completely suppressed. Furthermore, according to the Floquet expansion, when $\varphi_x = \pi / 2$ or $3 \pi / 2$ and $\varphi_y = 0$ or $\pi$, $\hat{H}^{(n)}$ is equal to $\hat{H}^{(- n)}$ in Eq.~\eqref{eq:DD_Cl_H_r1}, thereby inhibiting the first order of noises about $b_y, b_z$. The first order of noises about $b_x$ can also be inhibited by alternatively tuning $\varphi_y = 0$ or $\varphi_y = \pi$ for two nearest neighboring intervals and ensuring $|\gamma| T = 2 n \pi$ in each interval~\cite{note3}. Until now, we have demonstrated the complete suppression of both the zeroth and first orders of stray magnetic noises. Besides, higher orders of noises are also slightly suppressed by small coefficients because of the high-order Bessel function. More significantly, the above discussions about classical magnetic fields are also valid for quantum noises, only regarding $\bm{b}$ as $\hat{\bm{b}}$ operator. In the following two sections, we numerically calculate two concrete examples to illustrate that our protocol is capable of suppressing both classical and quantum noises.

\begin{figure}[t]
    \centering
    \includegraphics[width=7.8cm]{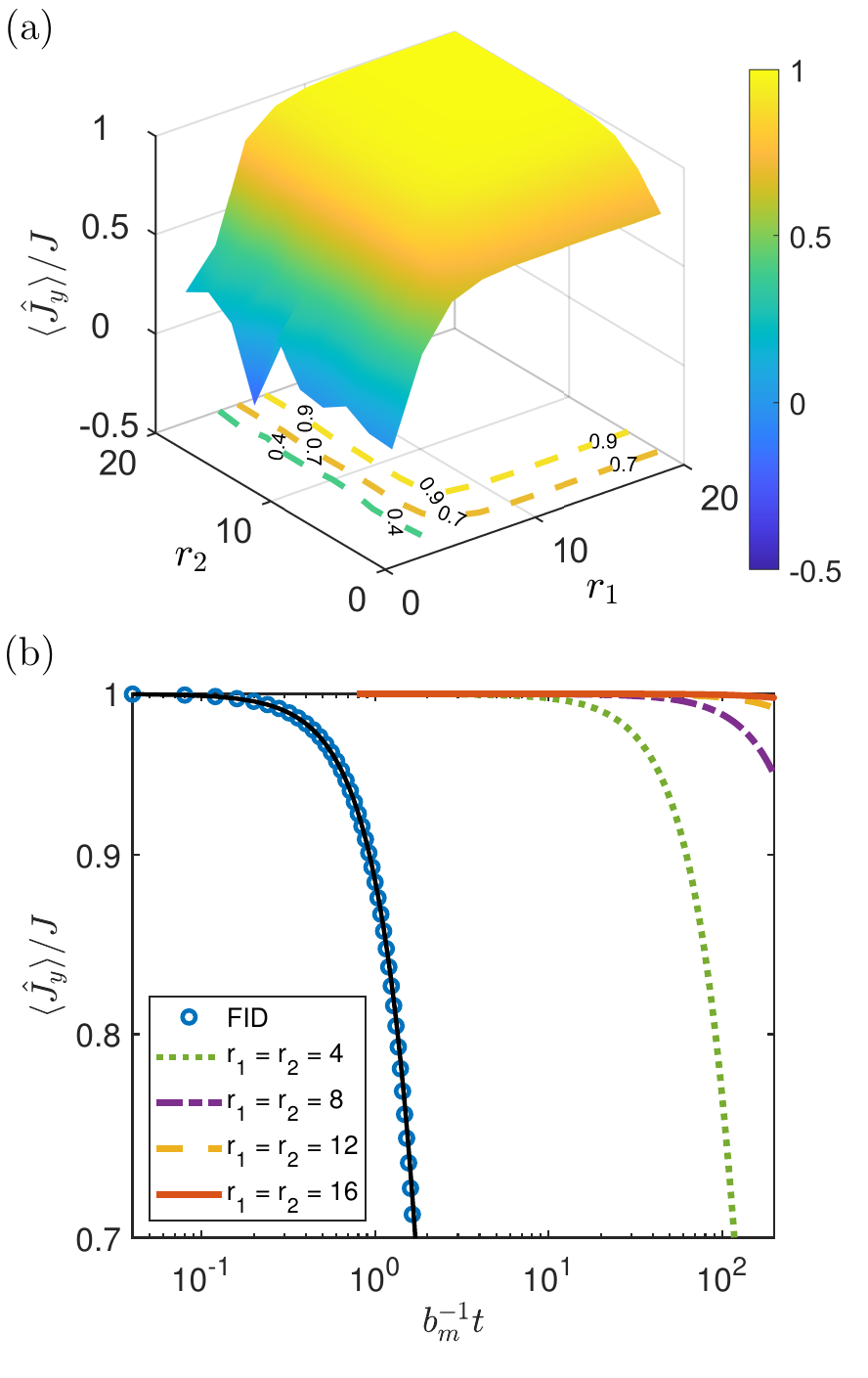}
    \caption{Suppression of classical noises with Floquet DD in an atomic spinor BEC. (a) Expectation values of $\hat{J}_y$ at time $b_m^{- 1} t = 200$ under different controlling parameters, $r_1 = \omega / \gamma$ and $r_2 = \gamma / b_m$. (b) Dynamics of $\left\langle \hat{J}_y \right\rangle$ under several typical pairs of $\{r_1, r_2\}$ and FID means there is no control. The black solid line mimics the decaying dynamics by $e^{- \sigma^2 t^2}$ with $\sigma \approx 0.35$. $\omega_x = \omega_y = \omega$, $\Omega_x = \omega r_0$, $\Omega_y = \gamma / \mathcal{J}_1(r_0)$, with $\mathcal{J}_0(r_0) = 0$. Stray magnetic fields $\bm{b}$ are randomly distributed in a sphere and the radius of the sphere is $b_m$. The initial state is polarized along $y$ direction. $J$ is the mean value of $\hat{J}_y$ for the initial state.}
    \label{fig:DD_Cl}
\end{figure}

\begin{figure}[t]
    \centering
    \includegraphics[width=7.8cm]{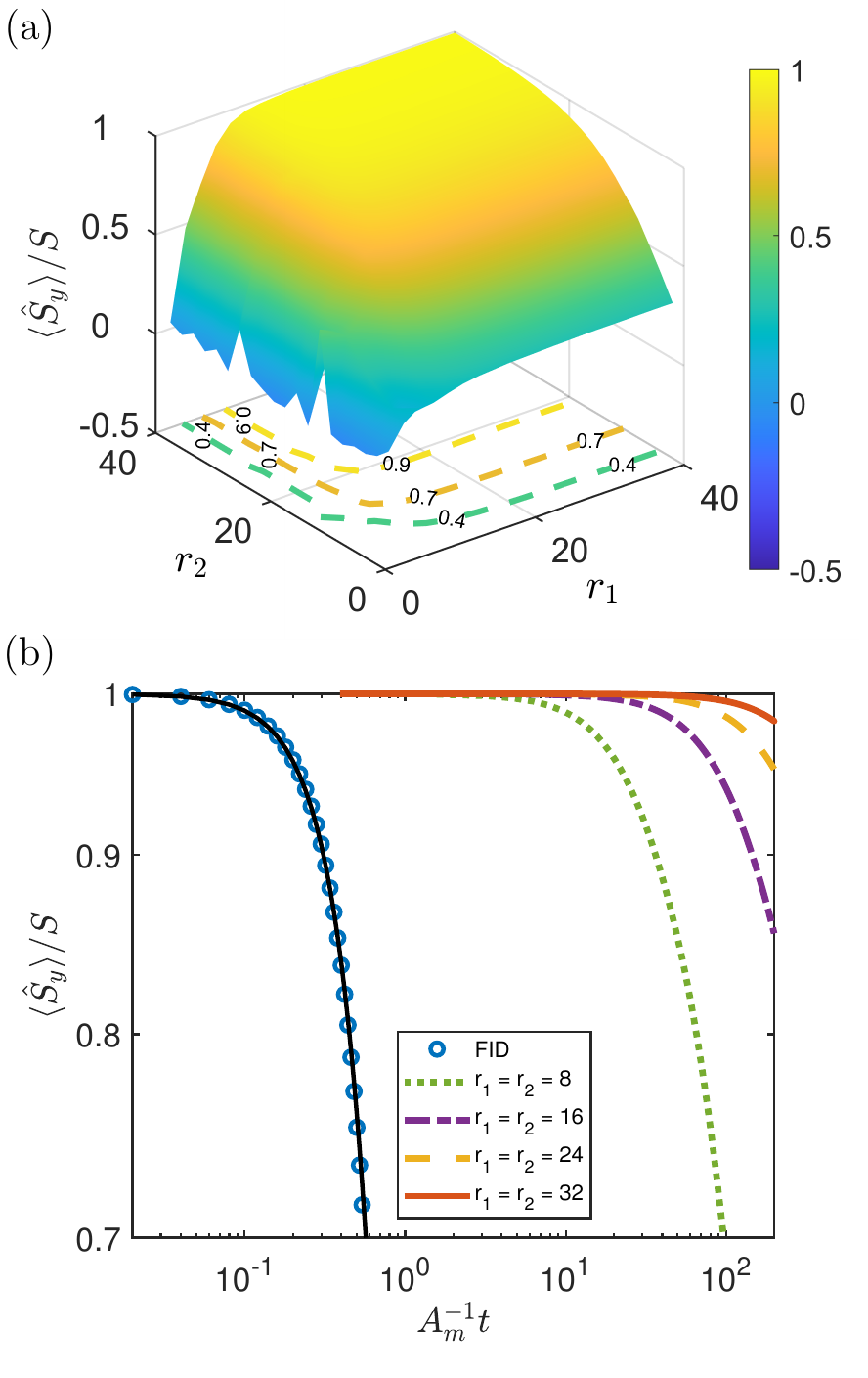}
    \caption{Suppression of quantum noises with Floquet DD in a GaAs QD. (a) Expectation values of $\hat{S}_y$ at time $A_m^{- 1} t = 200$ under different controlling parameters, $r_1 = \omega / \gamma$ and $r_2 = \gamma / b_m$. (b) Dynamics of $\langle \hat{S}_y \rangle$ under several typical pairs of $\{r_1, r_2\}$ and FID means there is no control. The black solid line mimics the decaying dynamics by $e^{- \sigma^2 t^2}$ with $\sigma \approx 0.35 \times 3$. $\omega_y = \omega_x = \omega$, $\Omega_x = \omega r_0$, $\Omega_y = \gamma / \mathcal{J}_1(r_0)$, with $\mathcal{J}_0(r_0) = 0$. Coupling strengths of nuclear spin interactions are distributed in a 2D (4 $\times$ 3) Gaussian form, $A_k = A_m \text{exp}\left[ - (x - x_0)^2 / w_x^2 - (y - y_0)^2 / w_y^2 \right]$, with effective widths $w_x / a_x = 3 / 2$ and $w_y / a_y = 2$ and a shifted center $x_0 / a_x = 0.1$ and $y_0 / a_y = 0.2$. The nuclear spins are initially randomly polarized on the Bloch sphere. The initial electronic spin is polarized along $y$ direction. $S$ is the mean value of $\hat{S}_y$ for the initial state.}
    \label{fig:DD_Qu}
\end{figure}

{\it{Suppressing classical noises in a spinor BEC.}} As for classical noises, we consider an atomic spinor BEC in external stray magnetic fields. The system with controls is described by the Hamiltonian $\hat{H} = \bm{b \cdot \hat{J}} + \sum_{\alpha = x, y} \Omega_\alpha \cos(\omega_\alpha t + \varphi_\alpha) \hat{J}_\alpha$ under the single-mode approximation~\cite{Ho1998Spinor, Ohmi1998Bose, Law1998Quantum, Stamper2013Spinor}, which is the same as Eq.~\eqref{eq:DD_Cl_H}. $\bm{\hat{J}}$ is the total spin of the spinor BEC and $\bm{b}$ are stray magnetic fields that are randomly distributed in a sphere with radius $b_m$. The controls are chosen along $x$ and $y$ axes. The energy and the time units are $b_m$ and $b_m^{- 1}$, respectively. 

As we have demonstrated based on the Floquet theory, stray magnetic fields $b_x, b_y, b_z$ are all suppressed to the second order. So without loss of generality, we just show numerical simulations of $\langle \hat{J}_y \rangle$ in Fig.~\ref{fig:DD_Cl}. Due to $\mathcal{J}_0(\Omega_x / \omega) = 0$, we set $r_0 \equiv  \Omega_x / \omega \approx 2.4048$. Besides, to satisfy the high-frequency expansion ($\omega > \Omega_y \simeq \gamma > b_m$) and the commensuration between two rotating wave transformations, we set $r_1 \equiv \omega / \gamma \in \mathbb{Z}, r_2 \equiv \gamma / b_m \in \mathbb{R}$. According to numerical results shown in Fig.~\ref{fig:DD_Cl} (a), we find expectation values of $\hat{J}_y$ approach to $J$ as $r_1$ and $r_2$ increase. In Fig.~\ref{fig:DD_Cl} (b), we show five typical dynamics of $\langle \hat{J}_y \rangle$ under different controlling parameters, where blue circles depict the free induced decay (FID) without Floquet DD, and other colored lines show decoherence with Floquet DD. These numerical results indicate that the coherence time is significantly prolonged under controls even when the controlling strengths are slightly larger than the strength of noises. For example, when $r_1 = r_2 = 4$, the coherence time is dozens of times longer than that of FID based on the rough estimation, and when $r_1 = r_2 \geq 8$, the coherence time is clearly seen prolonged by 2 orders of magnitude, which are consistent with the theoretical analysis indicating that stray magnetic fields are suppressed to the second order $O(1 / r^2_{1, 2})$.

{\it{Suppressing quantum noises in a QD.}} As for quantum noises, which are intrinsically different from classical noises described by stochastic fields, they are treated by operators and their correlations. To illustrate the suppressing capability of our protocol, we consider a gate-defined GaAs semiconductor QD system, which is well described by a central electron spin decohered by surrounding nuclear spins~\cite{Petta2005Coherent, Koppens2006Driven, Taylor2007Relaxation}. The Hamiltonian of this system with $N$ nuclear spins under two sequences of alternative controls is $\hat{H} = \bm{S} \cdot \sum_{k = 1}^N A_k \bm{I}_k + \sum_{\alpha = x, y} \Omega_\alpha \cos(\omega_\alpha t + \varphi_\alpha) \hat{S}_\alpha$, where $\bm{S}$ and $\bm{I}$ are the electron spin and the nuclear spin, respectively, and both are assumed spin-$1 / 2$ for simplicity. In general, $A_k$ is proportional to the local density of the electron at the position of the $k$-th nucleon, $A_k = A_m \text{exp}\left[ - (x - x_0)^2 / w_x^2 - (y - y_0)^2 / w_y^2 \right]$, a 2D Gaussian form with effective widths $w_x / a_x = 3 / 2$ and $w_y / a_y = 2$ and a shifted center $x_0 / a_x = 0.1$ and $y_0 / a_y = 0.2$~\cite{Dobrovitski2006Long}. The energy and the time units are $A_m$ and $A_m^{- 1}$. Here we neglect the interactions between nuclear spins because they are so small that the timescale of their dynamics is much longer than the decoherence time of the electron spin.

\begin{figure}[t]
    \centering
    \includegraphics[width=7.8cm]{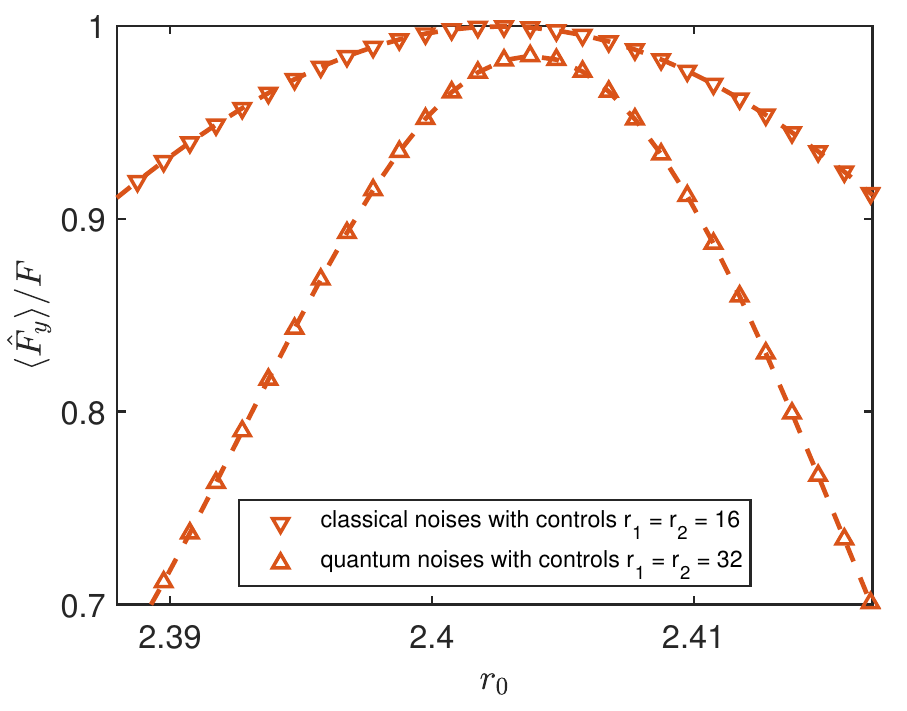}
    \caption{Robustness of Floquet DD under fluctuation in controlling powers. $\langle \hat{F}_y \rangle$ means the expectation value of spins along $y$-axis at time $b_m^{- 1} t = 200$ ($A_m^{- 1} t = 200$). $F$ denotes $J$ ($S$) for classical (quantum) noises. $r_0 \in [2.388, 2.417]$ corresponds to $1.2\%$ fluctuations of controlling power $\Omega_\alpha$. We set $r_1 = r_2 = 16$ and $r_1 = r_2 = 32$ for suppressing classical noises and quantum noises, respectively. The down (up) triangles depict the results for suppression of classical (quantum) noises with fluctuation of controlling power.}
    \label{fig:DD_Fluc}
\end{figure}

The meanings of controlling parameters $r_0$, $r_1$ and $r_2$ are the same as that in suppressing classical noises. We choose expectation values of $\hat{S}_y$ as the witness for quantum noises suppression. The system evolution is simulated by the Chebyshev polynomial expansion~\cite{Dobrovitski2003Efficient}. According to Fig.~\ref{fig:DD_Qu} (a), we find suppressing effects are increased as strengths of controlling parameters increase, which is similar to the behaviors in suppressing classical noises. However, based on Fig.~\ref{fig:DD_Qu} (b), we find that the coherence time for suppressing quantum noises at $r_1 = r_2 = 8$ is similar to that for suppressing classical noises at $r_1 = r_2 = 4$. To compare these two different types of noises, we use the decaying dynamics $e^{- \sigma^2 t^2}$ to mimic the processes of FID. We obtain $\sigma \approx 0.35$ for the classical noises, while $\sigma \approx 0.35 \times 3$ for the quantum noises. So in our numerical simulations, the controlling parameters for suppressing quantum noises should be larger than that for suppressing classical noises to achieve a similar coherence time.

{\it{Robustness of Floquet DD.}} Although we have demonstrated the ability of Floquet DD in suppressing both classical and quantum noises, controlling fluctuations arise from power, frequency and phase have not been considered. Based on current experimental conditions, we here mainly consider the fluctuation in controlling power $\Omega_{\alpha}$. Therefore, after applying the first rotating wave approximation the Hamiltonian in Eq.~\eqref{eq:DD_Cl_H} becomes $\hat{H}_{\text{eff}}^{\text{I}} = b_x \hat{J}_x + b'_y \hat{J}_y + b'_z \hat{J}_z + \gamma' \hat{J}_z$ with $b'_y = b_y \mathcal{J}_0(\Omega'_x / \omega)$, $b'_z = b_z \mathcal{J}_0(\Omega'_x / \omega)$, $\gamma' = - \Omega'_y \mathcal{J}_1(\Omega'_y / \omega) \sin(\varphi_x - \varphi_y)$. The zeroth order of noises about $b_y$ and $b_z$ is suppressed to $b'_y$ and $b'_z$, respectively. Then after applying the second rotating wave approximation, the above effective Hamiltonian becomes $\hat{H}_{\text{eff}}^{\text{II}} = b'_z \hat{J}_z - \left[ \Omega'_y \mathcal{J}_1(\Omega'_y / \omega) - \Omega_y \mathcal{J}_1(\Omega_y / \omega) \right] \sin(\varphi_x - \varphi_y)$. Additionally, the second term $- \left[ \Omega'_y \mathcal{J}_1(\Omega'_y / \omega) - \Omega_y \mathcal{J}_1(\Omega'_y / \omega) \right] \sin(\varphi_x - \varphi_y)$ can be cancelled by alternatively tuning $\varphi_y = 0$ or $\varphi_y = \pi$ for two nearest neighboring intervals. Finally, the fluctuation of $\Omega_\alpha$ leads to the residue of longitudinal magnetic noises $b'_z$, resulting in a shorter coherence time of spins in the $x$-$y$ plane compared to that along $z$-axis. Without loss of generality, we still choose $\langle \hat{F}_y \rangle$ as the witness of decoherence, where $\hat{F}$ denotes $\hat{J}$ for classical noises and $\hat{S}$ for quantum noises. Numerically, $r_0 \in [2.388, 2.417]$ approximately corresponds to $1.2\%$ fluctuations of $\Omega_\alpha$, and the numerical simulations are shown in Fig.~\ref{fig:DD_Fluc}. $\langle \hat{F}_y \rangle / F$ is still larger than $0.7$ at time $b_m^{- 1} t = 200$ and $A_m^{- 1} t = 200$ for classical noises and quantum noises, respectively. Based on these results, we find these two kinds of noises are suppressed well, and the coherence time is still prolonged by $2$ orders of magnitude even with fluctuation in controlling powers, which manifests that our protocol is robust against controlling fluctuations.

{\it{Conclusions and outlooks.}} We propose a Floquet DD protocol at zero bias to prevent systems from environmental noises. According to the theoretical analysis based on Floquet expansion, we find our protocol can not only suppress the zeroth order of noises but suppress the first order of noises. Besides, we numerically calculate two systems including a spinor BEC in classical stray magnetic fields and a GaAs quantum dot electron spin coupled to quantum nuclear spins. The numerical results of these two systems show our protocol significantly prolongs the coherence time and is robust against fluctuations in controlling powers. Furthermore, our protocol can be used to sense noises with high frequency by detecting decoherence of systems. Our results suggest alternative low-cost and high-portable DD techniques without bias, which may find wide applications in quantum computing, quantum information processing, nuclear magnetic resonance, magnetic resonance imaging. 

{\it{Acknowledgements.}} We thank Wenxian Zhang for invaluable discussions and Qi Yao for helpful discussions. This project is supported by the National Natural Science Foundation of China (NSFC) under Grants No. 12204428.

\normalem

\begin{thebibliography}{60}
\bibitem [{not({\natexlab{a}})}]{note2}%
      \BibitemOpen
      \href@noop {} {}\bibinfo {note} {The controlling axes are
      chosen as $x, y$ explicitly in our paper, however, we can select any other
      two perpendicular axes. Besides, we set $\omega_x = \omega_y$, however,
      decoupling effects can be obtained by a similar procedure when $\omega_y = n
      \omega_x$ with $n \in \mathbb{Z}$.}\BibitemShut {Stop}%
\bibitem [{not({\natexlab{b}})}]{note3}%
      \BibitemOpen
      \href@noop {} {}\bibinfo {note} {In this condition, the
      unitary evlotion after one period including two nearest neighboring intervals
      becomes $\hat{U}_{\text{eff}}^{\text{II}} = e^{- i \gamma_2 T \hat{J}_z}
      \hat{\mathcal{T}} e^{- i \int_{T / 2}^{T} \hat{H}_r^{\text{II}, 2}(t) dt}
      e^{i \gamma_2 T \hat{J}_z / 2} e^{- i \gamma_1 T \hat{J}_z / 2} \times \\
      \hat{\mathcal{T}} e^{- i \int_{0}^{T / 2} \hat{H}_r^{\text{II}, 1}(t) dt}$,
      with $\gamma_1 = -\gamma_2$, $|\gamma_j| T = 2 n \pi$ and
      $\hat{H}_r^{\text{II}, j} = b_x [\cos(\gamma_j t) \hat{J}_x - \sin(\gamma_j
      t) \hat{J}_y]$. Then $\hat{U}_{\text{eff}}^{\text{II}} \approx e^{- i 2 b_x^2
      \hat{J}_z / \gamma_2 + O(b_x^2 / \gamma_2^2)} e^{- i 2 b_x^2 \hat{J}_z /
      \gamma_1 + O(b_x^2 / \gamma_1^2)} = e^{O(b_x^2 / \gamma_2^2)}$.}\BibitemShut
      {Stop}%
\end{thebibliography}
%

\end{document}